\documentclass[conference]{IEEEtran}
\usepackage{xcolor}
\usepackage{bm}
\usepackage{amsfonts}
\usepackage{mathrsfs}
\usepackage{amssymb}
\usepackage{amsmath}
\usepackage{graphicx} 

\begin{document}
\title{Graph Convolutional Neural Networks for Optimal Power Flow Locational Marginal Price}

\author{\IEEEauthorblockN{Adrian-Petru Surani}
\IEEEauthorblockA{Cornell College of Engineering\\
Cornell Tech\\
New York, NY 10044 \\
Email: as3259@cornell.edu}
\and
\IEEEauthorblockN{Rahul Sahetiya}
\IEEEauthorblockA{Cornell College of Engineering\\
Cornell Tech\\
New York, NY 10044 \\
Email: rs2248@cornell.edu}}

\maketitle

\begin{abstract}
The real-time electricity market with the integration of renewable energies and electric vehicles have been receiving significant attention recently.  So far most of the literature  addresses the optimal power flow (OPF) problem in the real-time electricity market context by iterative methods.  However, solving OPF problems  in real-time is  challenging due to the high computational complexity by the iterative methods. Motivated by this fact, in this paper, we propose a Chebyshev Graph Convolutional Neural Networks (ChebGCN) to improve the efficiency of integrating low-carbon energy sources into power grids and to address scalability and adaptivity of end-to-end existing OPF solutions. The proposed GCN method is capable to predict the optimal energy market marginal prices in real time.  Numerical analysis is used to benchmark the results and validate the improvement.

\end{abstract}

\IEEEpeerreviewmaketitle

\section{Introduction}
Electricity market pricing is one of the core tasks of operating power-grids, as the real-time market determines the incremental adjustment for the day-ahead by solving the optimal-power flow (OPF) problem \cite{cain2012history}. The real-time OPF for market pricing also ensures high efficiency and reliability of grid operations, especially in the case of low-carbon energy sources integration, including renewable energy generations and electric vehicle integration. As the amount of smart, flexible resources' participation in distribution system operations keeps increasing, market-based approaches are needed for economic efficiency. \cite{li2015market} A key component to enable power distribution markets and provide economic incentives to market participants is the distribution locational marginal price (DLMP).

Typically, solving  OPF problems by iterative methods incurs excessive computational complexity, limiting its applicability in large-scale power networks and real-time implementations. 
The locational marginal price (LMP) represents the marginal cost to serve one incremental unit of demand at a specific location in electric power networks, \cite{oren1995nodal,litvinov2004marginal}. Typically the LMP is calculated by solving the direct current optimal power flow (DC OPF) problem. Due to high voltage volatility and high power losses due to high R/X ratio, reactive power modeling is needed. As the DC OPF approach is unable to capture power losses and reactive powers in distribution systems, it is unsuitable for calculating the DLMP. 
A different approach has been proposed to calculate DLMPs in \cite{li2015market}, \cite{li2013distribution}, \cite{yuan2016novel} which uses the single-phase alternating current optimal power flow (AC OPF) problem for the balanced system. At the same time, as LMPs relate to the duality analysis for OPF, their dependence on grid topology has been recognized in \cite{jia2013impact}, \cite{geng2016learning}. 

To improve the high computation complexity of the accurate AC OPF problem, which is non-convex and non-linear, machine learning (ML) techniques have been proposed throughout existing literature: identifying active constraints \cite{misra2022learning}, finding a warm start for iterative OPF solutions \cite {baker2019learning}, or addressing the feasibility issue \cite{zamzam2020learning}. These models require an extensive off-line training of neural network (NN) models and as they rely on end-to-end NNs, also incur a high model and computation complexity for large-scale power grids. An additional cost is represented by the re-training of the models whenever the system inputs change as a result of frequently varying topologies.

To address these problems, we are motivated to utilize graph convolutional neural networks (GCN) to approximate the optimal  marginal prices. The proposed method considers the power system measurements as the low-pass graph signals, and derive the suitable Graph Shift Operator (GSO) to design GCN. The proposed method also designs the regulation terms for the feasibility of power flow constraints. 

The rest of the paper is organized as follows: The AC OPF formulation is introduced initially in Section II. In Section III, the proposed approach is discussed including various adaptations and optimizations such as convolution in Fourier domain to speed up computations. The case study results on a transmission system are given in Section IV, while Section V concludes the paper and discusses future research directions and exploratory questions.

\section{Problem Formulation}

The LMP at a certain location represents the incremental cost to supply an extra unit of load at this location. The typical approach is to use the DC OPF problem to determine the LMPs in the transmission level. The formulation of the considered AC OPF problem is:

\begin{equation}
    \begin{array}{lll}
         &  \min \limits_{P_{G, k}} & \sum\limits_{k \in \mathcal{N}_G} C_i (P_{ {G}, k})\\
         & \text{s.t.} & P_{G, i} - P_{L, i} = \Re \{ V_i \cdot (I_i)^* \}, i\in \mathcal{N} \\
         & & Q_{G, i} - Q_{L, i} = \Im \left \{ V_i \cdot (I_i)^*\right \},   i\in \mathcal{N} \\ 
         & & \underline{P}_{G, k} \leq P_{G, k} \leq \overline {P}_{G, k}, k\in \mathcal{N}_G \\ 
         & & \underline{Q}_{G, k} \leq Q_{G, k} \leq \overline{Q}_{G, k}, k\in \mathcal{N}_G \\
         & & \underline {V_i} \leq |V_i| \leq \overline {V_i}, i\in \mathcal{N}
    \end{array}
    \label{eqn:opf}
\end{equation}
where $\mathcal{N}_G$ denotes the set of buses which have generators. $\Re\{ \cdot \}$ and $\Im\{ \cdot \}$ denote the real and the imaginary part of a complex number, respectively. $P_{G, k}$ and $Q_{G, k}$ represent the active and reactive power output of the generator at bus k. Similarly, $\underline{P}_{G, k}, \underline{Q}_{G, k}$ and $\overline{P}_{G, k}, \overline{Q}_{G, k}$ correspond to the lower bounds and upper bounds for the active and reactive power generation. $|V_i|$ corresponds to the voltage magnitude at bus $i$, and $\underline {V_i}, \overline {V_i}$ the associated lower and upper bounds.

For simplicity, let $u$ denote the vector of all control variables and $x$ the vector of all state variables, including the voltage magnitude and angle at every bus and phase. We define $h(x, u) \leq 0$ to represent the inequality constraints in \eqref{eqn:opf}. The Lagrangian function of the formulated AC OPF problem can be written as:

\begin{equation}
    \begin{array}{ll}\label{eqn:lag}
     &  L(x, u, \lambda, \nu, \mu) = \sum\limits_{k \in \mathcal {N}_\mathcal{G}} C_i (P_{G, k}) \\
     & - \sum_{i \in \mathcal{N}} \lambda_i \left ( P_{G, i} - P_{L, i} - \Re \{ V_i \cdot (I_i)^* \} \right ) \\ 
     & - \sum_{i \in \mathcal {N}} \nu_i \cdot (Q_{G, i} - Q_{L, i} - \Im \{   V_i \cdot (I_i)\}^* ) \\
     & + \sum\limits_{m \in \mathcal {H}} \mu_m \cdot h_m(x, u)
    \end{array}
\end{equation}
where $\mathcal{N}$ represents the set of all buses in the system, $\mathcal{P}_i$ the set of phases at bus i, and $\mathcal{H}$ the set of inequality constraints. $\lambda_i$ and $\nu_i$ are the Lagrange multipliers corresponding to the active power balance equation, and the reactive power balance equation.  $u_m$ is the Lagrange multiplier associated with the inequality constraint $h_m(x, u) \leq 0$.

Assuming the considered AC OPF problem has an optimal solution $(x^*, u^*)$, it can be calculated as:

\begin{equation}
    DLMP_{i} = \left . \frac {\partial f} {\partial P_{L, i}} \right | _{x^*, u^*} = \left . \frac {\partial L} {\partial P_{L, i}} \right | _{x^*, u^*} = \lambda_i\label{eqn:dual}
\end{equation}
Note that each bus has a marginal price $\lambda_i$.




\section{Proposed Approach}

The GCNs are used to learn a mapping function from voltage  measurements to optimal marginal prices. 

\subsection{Chebyshev Graph Neural Networks}

In spectral graph analysis, the graph Laplacian is an essential operator for undirected connected graphs, defined as: $\bm{L} = \bm{D} - \bm{A} \in \mathbb{R}^{{N} \times {N}}$, while the normalized Laplacian is $\bm{L} = \bm{I}_{ {N}} - \bm{D}^{- 1/2}\bm{AD}^{-1/2}$, where $\bm{I}_{{N}}$ is the identity matrix, $\bm{A}$ is the adjacent matrix, and $\bm{D} \in \mathbb{R}^{{N} \times {N}}$ is the diagonal degree matrix with $\bm{D}_{ii} = \sum_j \bm{A}_{ij}$. The Laplacian matrix is diagonalized by the Fourier basis $U = [u_0, ..., u_{{N} - 1}]$ $\in \mathbb {R}^{{N} \times {N}}$, where $\{u_l\}_{l = 0}^{\mathbb{N} - 1} \in \mathbb {R}^{{N}}$ are the orthogonal eigenvectors of $\bm{L}$. The eigenvalue decomposition for the Laplacian matrix is $L = \bm{U} \Lambda \bm{U}^T $, where $\Lambda = diag([\lambda_0, ... \lambda_{{N} - 1}]) \in \mathbb {R}^{{N} \times {N}}$ and $\{\lambda_l\}_{l = 0}^{\mathcal{N} - 1} \in \mathbb{R}^{\mathcal{N}}$ is the ordered list of non-negative real eigenvalues associated with $\bm{L}$, referred to as the graph's frequencies. The  voltage phasor at time $t$ is represented as the signal $x =  \Big[|\bm{V}|, \bm{\theta}\Big]  \in \mathbb {R}^{{2N} }$ and the graph Fourier transform of the signal $x \in \mathbb {R}^{{2N}}$ is $\hat{x} = \bm{U}^T x \in \mathbb {R}^{{2N}}$, and its inverse Fourier transform as $x = \bm {U}\hat{x}$. In the Fourier domain, the convolution on graph $\mathcal {G}$ is denoted $x_{*\mathcal{G}y} = U((U^T x) \odot (U^T y))$, where the $\odot$ represents the element-wise product. Filtering the signal $x$ by $g_\theta$:

\begin{equation}
    y = g_\theta (\bm {L}) x = g_\theta (\bm{U} \Lambda \bm{U}^T)x = \bm {U} g_\theta (\Lambda) \bm{U}^{T}x
    \label{eqn:filteredx}
\end{equation}

The ChebNet model uses mainly Chebyshev polynomials to replace the convolution kernel in the spectral domain to represent the filter, as defined in \ref{eqn:expansionfilter}, which is the truncated expansion of order $K - 1$:

\begin{equation}
    g_{\theta} (\Lambda) = \sum_{k = 0}^{K - 1} \theta_k T_k (\tilde{\Lambda})
    \label{eqn:expansionfilter}
\end{equation}

In   \eqref{eqn:expansionfilter}, the parameter $\theta \in \mathbb {R}^{{K}}$ is the vector of Chebyshev coefficients. $T_k(\tilde {\bm{\Lambda}}) \in \mathbb {R}^{\mathcal{N} \times \mathcal{N}}$ is the Chebyshev polynomial with order $k$ evaluated at $\tilde{\bm \Lambda} = 2 \lambda / \lambda_{max} - \bm{I}_{\mathcal{N}}$, where $\tilde{\Lambda}_{ii} \in [-1, 1]$ and $\lambda_{max}$ is the maximum eigenvalue.

The calculation of the Chebyshev polynomial, $T_k(x)$ can be obtained through the following recursive equation:
\begin{equation}
\left\{\begin{array}{lll}
    T_0(x) = 1 \\
    T_1(x) = x \\
    T_{\mathcal{N} + 1}(x) = 2xT_{\mathcal{N}}(x) - T_{\mathcal{N} - 1}(x) 
  \end{array}\right.
\end{equation}

Therefore, the convolution operation is expressed as:
\begin{equation}
    x_{*\mathcal{G}g_{\theta}} = \bm {U} (\sum_{i = 0}^{K - 1} \theta_k T_k (\tilde{\bm {\Lambda}})) \bm {U}^T x
\end{equation}
Similar to $\tilde {\bm{\Lambda}} = 2 \bm {\Lambda} / \lambda_{max} - \bm {I}_{\mathcal {N}}$ and $T_k (\tilde {\Lambda})$, $\bm {\tilde{L}}$, and $T_i (\bm {\tilde{L}})$ can be defined as: 
\begin{equation}
    \tilde {\bm {L}} = 2 \bm{L} / \lambda_{max} - \bm {I}_{\mathcal{N}}
\end{equation}
and
\begin{equation}
    \begin{array}{ll}
        T_k (\tilde{\bm {L}}) &= T_k (2 \bm{L} / \lambda_{max} - \bm {I}_{\mathcal{N}}) \\
         &= T_k (2 \bm{U \Lambda U^T} / \lambda_{max} - \bm {UI}_{\mathcal{N}}U^T) \\
         &= T_k (\bm{U} (2 \bm{\Lambda} / \lambda_{max} - \bm {I}_{\mathcal{N}})U^T) \\
         &= \bm{U} T_k (\tilde {\bm {\Lambda}}) \bm {U}^T
    \end{array}
\end{equation}

By performing the Fourier transform, doing the convolution and then recovering the $\bm{U}$ through the inverse Fourier transform, we can avoid extra-calculations, and simplify the equation further:
\begin{equation}
    \begin{array}{ll}
            x_{*\mathcal{G}g_{\theta}} & = \bm {U} \left ( \sum \limits_{i = 0}^{K - 1} \theta_k T_k (\tilde {\bm {\Lambda}}) \right ) \bm {U}^T x \\
         & = \sum \limits_{i = 0}^{K - 1} \theta_k \left ( \bm{U} T_k (\tilde {\bm {\Lambda}})\bm{U}^T \right ) x \\
         & = \sum \limits_{i = 0}^{K - 1} \theta_k T_k \left ( \tilde {\bm {{L}}} \right ) x
    \end{array}
    \label{eqn:ftsimplification}
\end{equation}

As pointed above in \eqref{eqn:ftsimplification}, the calculation does not require eigenvector decomposition, which improves the computational complexity.
Given a sample   $x$ , we calculate the output  output feature map denoted as $y$, using the $\theta_{i, j} \in \mathbb{R}^K$ as Chebyshev coefficients.
\begin{equation}
    y = \sum \limits_{k = 1}^{K} g_{\theta_{k}} (\bm {L})x
\end{equation}
where the output  can be multiple channels.

\subsection{GCN for Marginal Price Prediction and Forecasting}
Inspired by the graph signal   $x$    in \eqref{eqn:ftsimplification}, we propose to utilize Chebshev GCN to predict the marginal prices, i.e., $\lambda_i$ of \eqref{eqn:dual},  instead of solving a nonconvex optimization problem. In our problem, our input graph signals are the voltage magnitude and voltage angles. Our graph shift operator $\bm{L}$ is the absolute value of system matrix $\bm{Y}$, i.e., $\bm{L} = |\bm{Y}| $. The process of training and testing processes can be summarized as follows.
\begin{enumerate}
    \item Utilize the Interior Point Method (IPM) to solve the Lagragian problem in \eqref{eqn:lag} to obtain the optimal $\lambda_i^*, i\in \mathcal{N}$.
    \item Train the graph neural networks by regarding $|V_i|, i\in \mathcal{N}$ and $|\theta_i|, i\in \mathcal{N}$ as inputs, and $\lambda_i, i\in \mathcal{N}$ as the outputs.
    \item After training the GCN, we can predict the unseen $\lambda_i, i\in \mathcal{N}$ by putting the new-arrival measurements  $|V_i|, i\in \mathcal{N}$ and $|\theta_i|, i\in \mathcal{N}$.
\end{enumerate}
This framework can be also implemented  for the forecasting problem if the power demands are time-series measurements. The only difference is Step 2, i.e., regarding $|V_i(t)|, i\in \mathcal{N}$ and $|\theta_i(t)|, i\in \mathcal{N}$ as inputs, and the future $\lambda_i(t+1), i\in \mathcal{N}$ as the outputs.

\section{Analysis and numerical results}
We will consider the IEEE 118-bus standard testing cases with the real-world renewable energies and EV charging powers for analysis. The 118-bus system has been used to represent the real-world demand, and for each demand vector the real and imaginary parts were separated.
 
\subsection{System Setting}


The code was run on a machine powered by the Apple M1 chip, using 16 GB RAM. Most of the computations were run under an upper bound of 2 hours. The Python environment was built using the Conda package system and Python version 3.10.8. Pytorch 1.13.0 was used to build the GNN models and to subsequently benchmark them. We consider two benchmarks, i.e., fully connected neural networks (FCNN) and 1st-order approximated graph neural network \cite{kipf2016semi}.

\subsection{Prediction of Marginal Prices}

\begin{figure}[!htp]
  \center
    \includegraphics[width=0.42\textwidth]{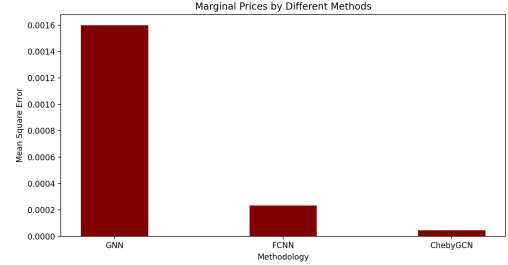}
  \caption{The MSE comparison between Chebyshev GCN with different methods.}\label{Plotbar_MSE}
  \vspace{-0.6cm}
\end{figure}

\begin{figure}[!htp]
  \center
    \includegraphics[width=0.42\textwidth]{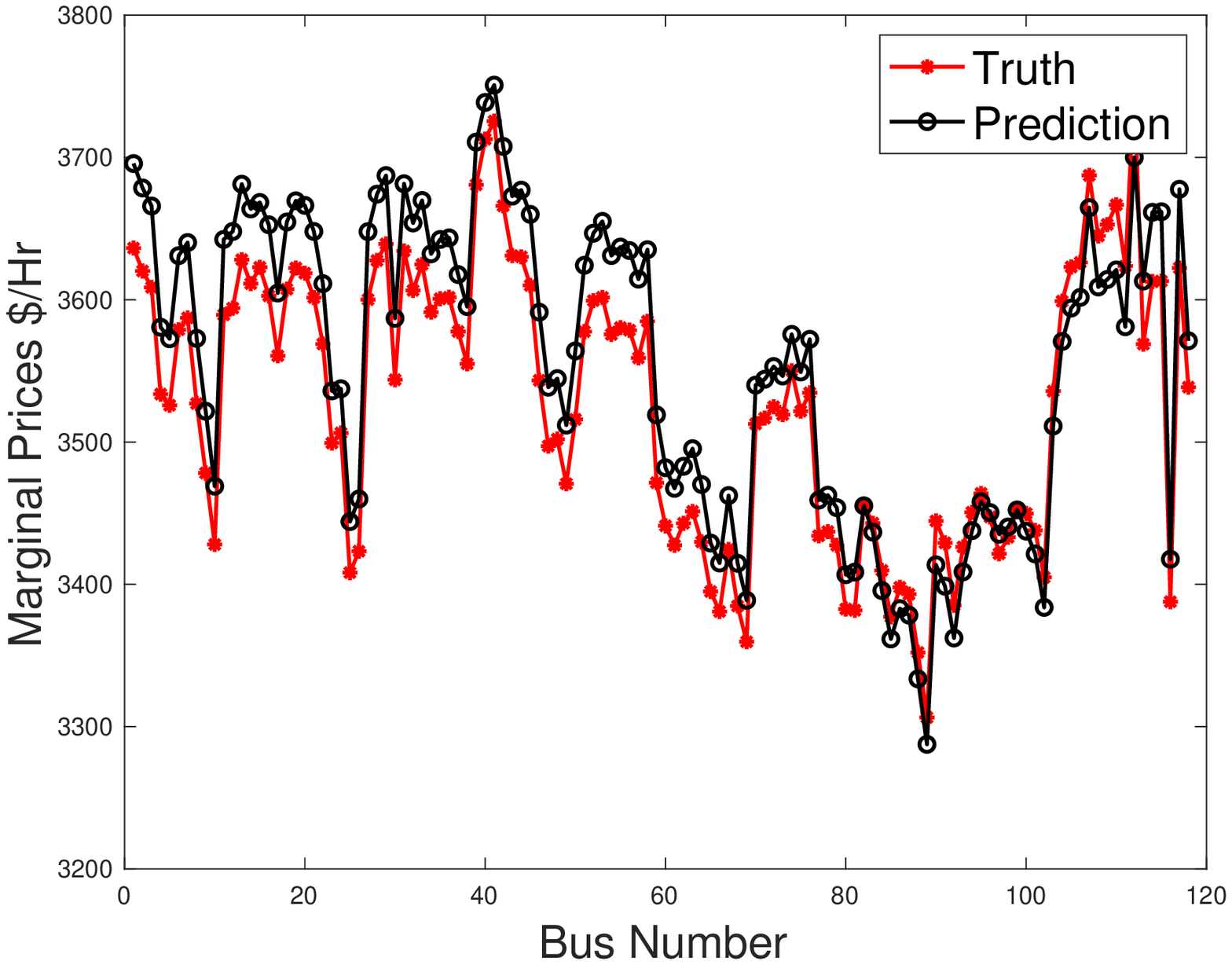}
  \caption{The marginal price prediction by the FCNN method.}\label{FCNN}
    \vspace{-0.6cm}
\end{figure}

\begin{figure}[!htp]
  \center
    \includegraphics[width=0.42\textwidth]{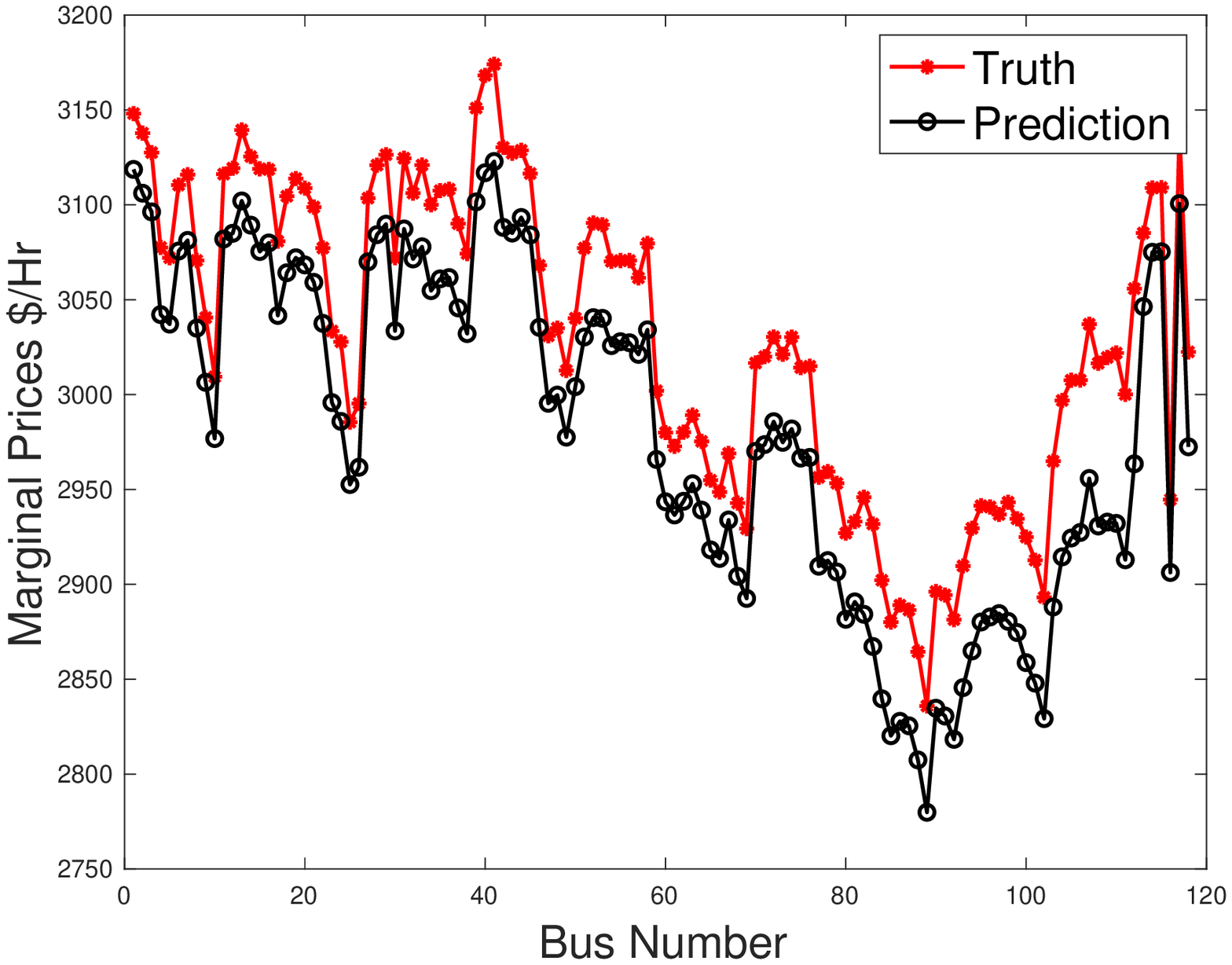}
  \caption{The marginal price prediction by the GNN method.}\label{GNN}
      \vspace{-0.6cm}
\end{figure}

\begin{figure}[!htp]
  \center
    \includegraphics[width=0.42\textwidth]{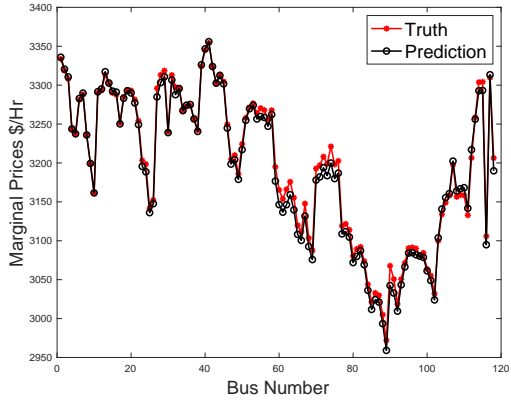}
  \caption{The marginal price prediction by the Chebyshev GCN method.}\label{ChebyGNN}
      \vspace{-0.6cm}
\end{figure}

In order to predict marginal prices, three approaches were used based on PyTorch's classic models: Fully-Connected Neural Networks (FCNNs), GNN (Graph Neural Networks) and Chebyshev GCN. 
Figure \ref{Plotbar_MSE} compares the performances of the three methods and subsequently the improvements throughout the stages. In particular, Chebyshev GCN predicts the optimal marginal prices with $4.5717e^{-5}$ MSEs, while GNN has 0.0016 and FCNN has $2.3304e^{-4}$ MSEs.
The GNN performance has been significantly better in both prediction and forecasting. We also illustrate the performance of FCNN, GNN, and Chebyshev GCN in  Figs. \ref{FCNN}, \ref{GNN} and \ref{ChebyGNN}. It shows that Chebyshev GCN can predict the ground-truth marginal prices with very high accuracy.

\subsection{Forecasting of Marginal Prices}
\begin{figure}[!htp]
  \center
    \includegraphics[width=0.42\textwidth]{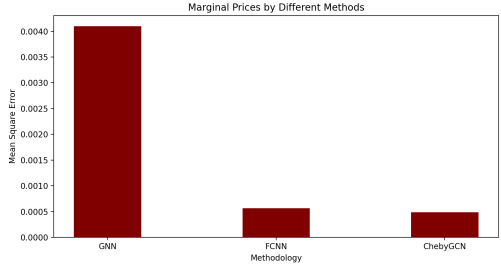}
  \caption{The MSE comparison between Chebyshev GCN with different methods.}\label{Plotbar_MSE_Pred}
\vspace{-0.5cm}
\end{figure}

\begin{figure}[!htp]
  \center
    \includegraphics[width=0.42\textwidth]{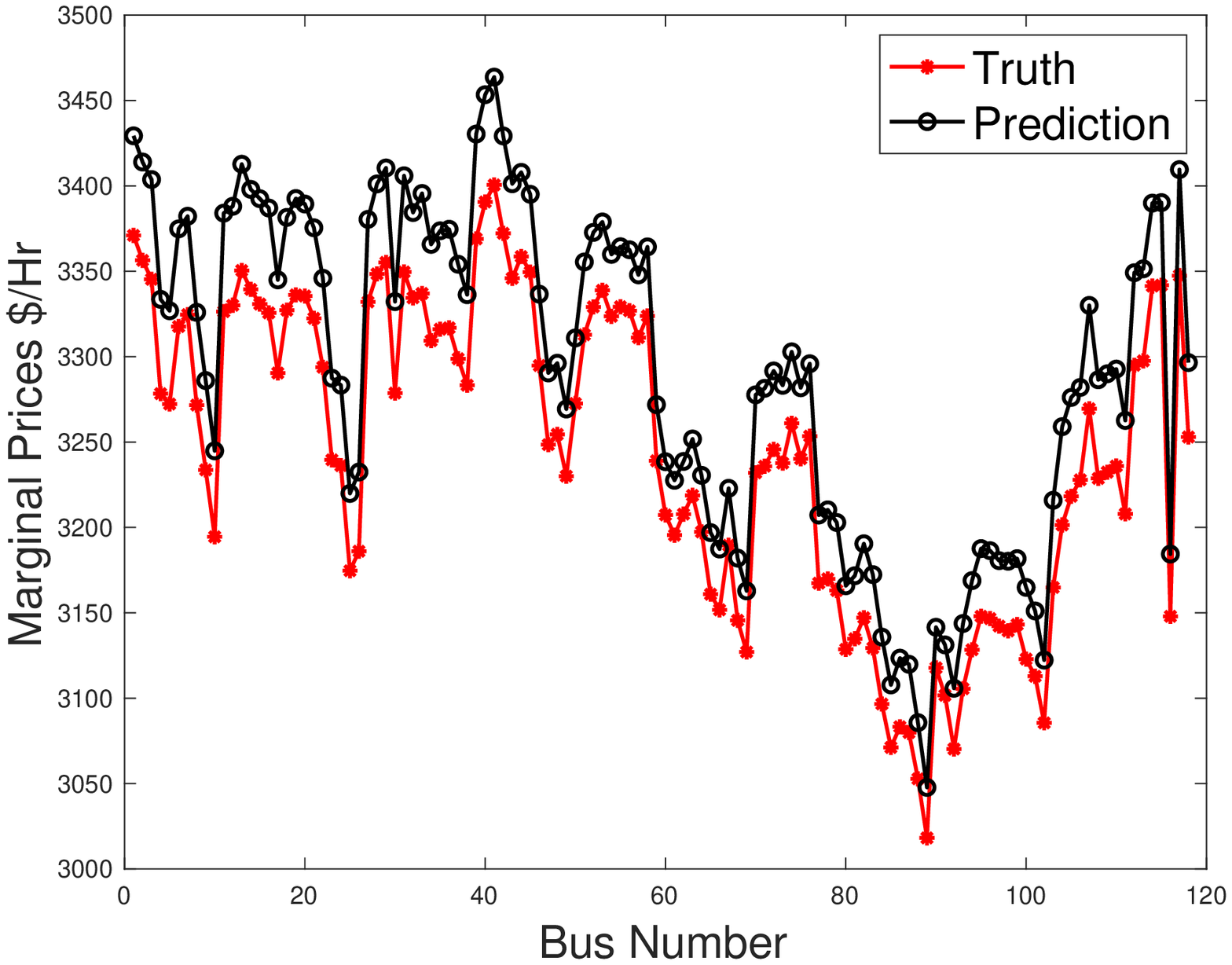}
  \caption{The marginal price forecasting by the FCNN method.}\label{FCNN_Pred}
  \vspace{-0.4cm}
\end{figure}

\begin{figure}[!htp]
  \center
    \includegraphics[width=0.42\textwidth]{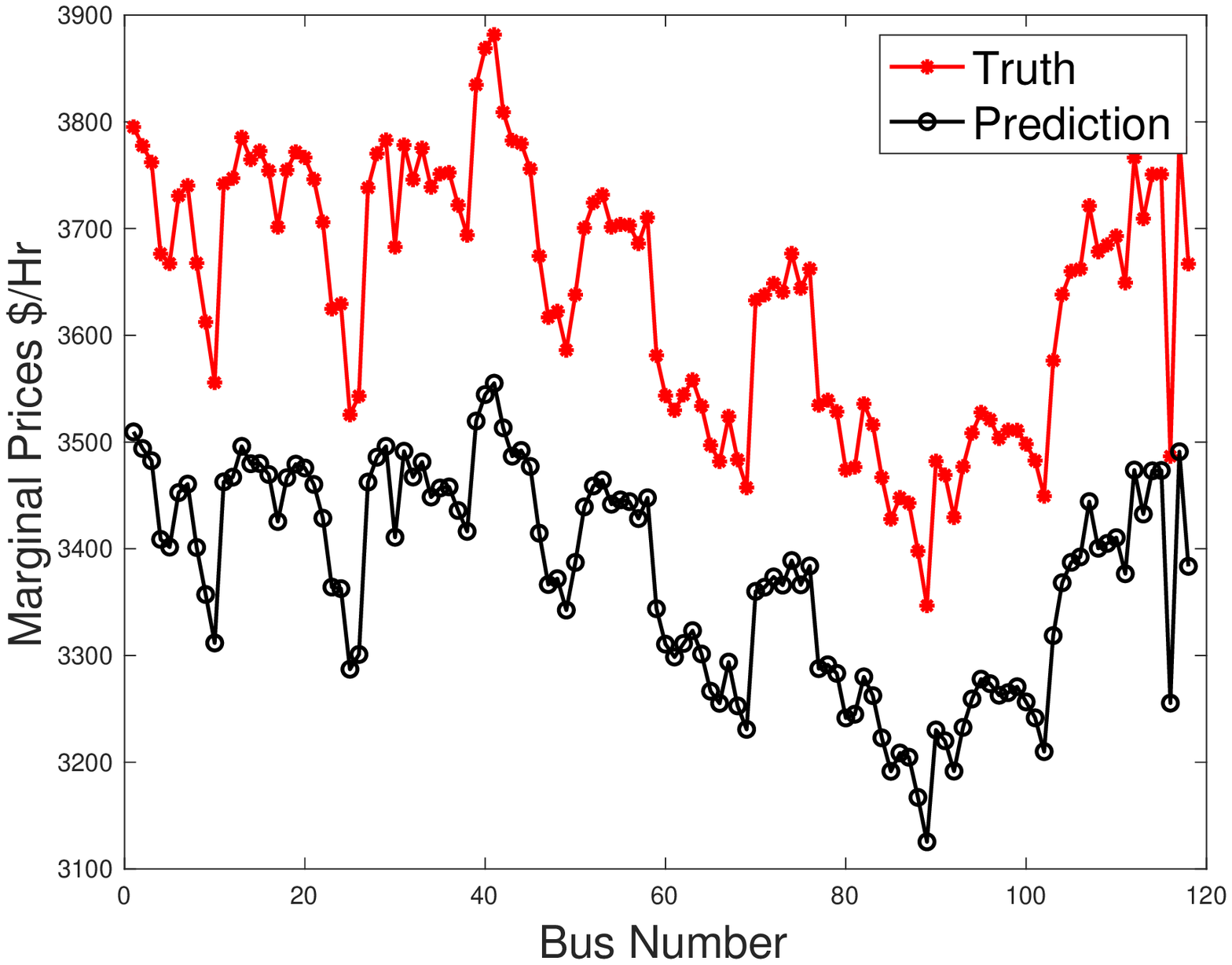}
  \caption{The marginal price forecasting by the GNN method.}\label{GNN_Pred}
  \vspace{-0.4cm}
\end{figure}

\begin{figure}[!htp]
  \center
    \includegraphics[width=0.42\textwidth]{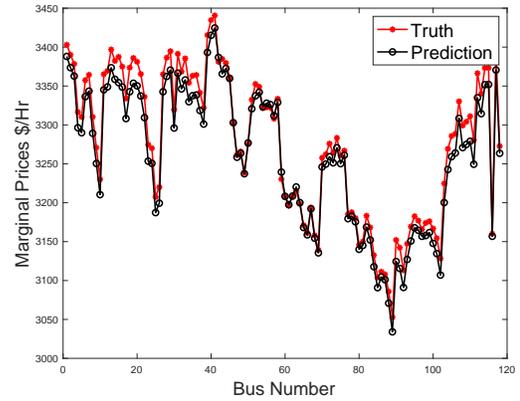}
  \caption{The marginal price forecasting by the Chebyshev GCN method.}\label{ChebyGNN_Pred}
\vspace{-0.4cm}
\end{figure}

We will also consider the forecasting of marginal prices for the future. Figure \ref{Plotbar_MSE_Pred} compares the performances of the three methods, i.e., FCNN, GNN, and Chebyshev GCN. Particually, GNN, FCNN and ChebyGCN forecast  the marginal prices with 0.0041, $5.6336e^{-4}$, and $4.8788e^{-4}$  MSEs. The results show that ChebyGCN performs slightly better than FCNN. This is because forecasting is a time-series problem, while ChebyGCN does not consider the temporal correlations. We also illustrate three examples of FCNN, GNN and Chebyshev GCN for the forecasting marginal prices in Figs. \ref{FCNN_Pred}-\ref{ChebyGNN_Pred}. The Chebyshev GCN still produces a promising result that approximates the ground-truth marginal prices.

\section{Conclusion}

In this paper, a graph convolutional neural network (GCN) method is used to approximate optimal marginal prices, based on the definition and further calculations of DLMPs. 

Future research directions to be considered can be separated in two areas: how market mechanisms can be established using the AC OPF based DLMP to provide right incentives, and how market participants react to such AC OPF based DLMPs in order to make optimal decisions. Both of these tracks can be investigated to answer questions regarding the way DLMPs lead to more efficient power delivery in transmission systems.

A secondary problem is based on the impact of reactive power on DLMPs, which is still under consideration. Studying voltage constraints and how they contribute to DLMPs, and how reactive power components interact with the phase imbalance and system losses would bring additional detail into the way methods described in the current work perform. Further case studies may also be needed to quantify the significance of each component in DLMPs. 

A separate extension to be considered is the inclusion of Electric Vehicle (EV) charging stations as sources for our graph. However, a different problem to be considered revolves around calculating the local marginal price (LMP) in EV charging settings, which brings additional complexity both in terms of larger number of constraints, and actual data generation. This last problem can also be solved for large-scale vehicles as a final extension.

Finally, better accuracy during the GNN predictions can be obtained by considering time-series data, which would require Recurrent Neural Networks (RNNs) and Long Short-Term Memory (LSTM).

\begin{footnotesize}

\end{footnotesize}

\end{document}